\documentclass[12pt]{article}
\let\section=\subsection     \let\subsection=\subsubsection                
\usepackage{graphicx}
\usepackage{epsfig}
\usepackage[tbtags]{amsmath}

\newcommand{\beq}{\begin{equation}}
\newcommand{\eeq}{\end{equation}}
\newcommand{\beqa}{\begin{eqnarray}}
\newcommand{\eeqa}{\end{eqnarray}}
\newcommand{\no}{\nonumber}

\newcommand{\sfrac}[2]{{\textstyle\frac{#1}{#2}}}
\newcommand{\mnod}{{\overset{\; \circ}{M}}}

\begin{document}
\begin{center}
   {\large \bf $\mbox{\boldmath$\eta, \eta'$}$ photoproduction and}\\[2mm]
   {\large \bf electroproduction off protons}\\[5mm]
   B.~Borasoy, E.~Marco, S.~Wetzel \\[5mm]
   {\small \it  Physik Department -T39, TU M\"unchen, \\
   85747 Garching, Germany \\[8mm] }
\end{center}

\begin{abstract}\noindent
Photo- and electroproduction of $\eta, \eta'$ mesons on protons
are investigated within a relativistic chiral unitary
approach based on coupled channels. The $s$ wave potentials
for electroproduction and meson-baryon scattering are derived from a chiral
effective Lagrangian which includes the $\eta'$ as an explicit degree of
freedom and incorporates important features of the underlying QCD
Lagrangian such as the axial $U(1)$ anomaly. The effective potentials
are iterated in a Bethe-Salpeter equation and cross sections for $\eta,
\eta'$ photo- and electroproduction from protons are obtained.
The results for the $\eta'$ photoproduction cross section
reproduce the appearance of an $S_{11}$ resonance
around 1.9 GeV observed at ELSA.
The inclusion of electromagnetic form factors increases 
the predicted $\eta$ electroproduction cross sections, 
providing a qualitative explanation
for the hard form factor of the photocoupling amplitude observed
at CLAS.
\end{abstract}

\section{Introduction}
Photoproduction of mesons is a tool to study baryon resonances and the
investigation of transitions between these states provides a crucial test for
hadron models. 
Because of their hadronic decay modes nucleon resonances have large overlapping
widths, which makes it difficult to study individual states, but selection
rules in certain decay channels can reduce the number of possible resonances.
The isoscalars $\eta$ and $\eta'$ are such examples since -- due to isospin
conservation -- only the isospin-$\frac{1}{2}$ excited states decay into the
$\eta N$ and $\eta' N$ channels.
Electroproduction experiments are even more sensitive to the structures of the nucleon
due to the longitudinal coupling of the virtual photon to the
nucleon spin and might in addition yield some insight into a possible
onset of perturbative QCD.

In this work, we restrict ourselves to the low-energy region where
non-perturbative QCD dominates.
Chiral symmetry is believed to govern interactions among hadrons at low energies
where the relevant degrees of freedom are not the quark and 
gluon fields of the QCD Lagrangian, but composite hadrons. In order to make contact
with experiment one must resort to non-perturbative methods such as chiral
effective field theory which incorporates the symmetries and
symmetry breaking patterns of the underlying theory QCD and is written in terms of the 
active degrees of freedom.
A systematic loop expansion can be carried out in the framework of chiral perturbation theory (ChPT)
which inherently involves a
characteristic scale $\Lambda_\chi = 4 \pi F_\pi \approx 1.2$ GeV at which the
chiral series is expected to break down. The limitation to very low-energy processes
is even enhanced in the vicinity of resonances. The appearance of resonances
in certain channels constitutes a major problem to the loopwise expansion of ChPT,
since their contribution cannot be reproduced at any given order of the chiral series.
Recently, considerable effort has been undertaken to combine the chiral effective
Lagrangian approach with 
the Bethe-Salpeter equation making it possible to go to energies beyond $\Lambda_\chi$ 
and to generate the resonances dynamically \cite{KSW}.
Two prominent examples of resonances in the baryonic sector are the $\Lambda(1405)$ and the
$S_{11}(1535)$.

The  $\eta'$ is closely related to the axial $U(1)$ anomaly. 
The QCD Lagrangian with massless quarks
exhibits an $SU(3)_L \times SU(3)_R$ chiral symmetry which is broken down
spontaneously to $SU(3)_V$, 
giving rise to a Goldstone boson octet of pseudoscalar mesons
which become massless in the chiral limit of zero quark masses.
On the other hand, the axial $U(1)$ symmetry of the QCD Lagrangian is broken by
the anomaly which manifests itself phenomenologically in the mass of the $\eta'(958)$ which
is considerably heavier than the masses of the (pseudo-) Goldstone bosons.

The experimental data for $\eta'$ photoproduction from ELSA
\cite{Pl} suggested the coherent excitation of two resonances 
$S_{11} (1897)$ and $P_{11} (1986)$.
In this work we will restrict ourselves to $s$ waves and therefore the
comparison with data should only be valid in the near threshold region.
One of the purposes of this work is to shed some light on the $s$ wave resonance $S_{11} (1897)$.
Furthermore, this investigation will provide a test whether processes up to
energies of $\sqrt{s} \sim 2$ GeV
are still constrained by chiral symmetry 
and whether the $\eta'$ meson can be included in the
effective Lagrangian with baryons as proposed in \cite{B}.

\section{Sketch of the calculation} \label{sec:scetch}

In this section we briefly outline the calculation. In the first part, features of
the effective Lagrangian are presented, while in the second part the coupled channel
approach is introduced and then generalized to photo- and electroproduction 
processes.

\subsection{The effective $\mbox{\boldmath$U(3)$}$ Lagrangian}  \label{sec:Lagran}

In the effective field theory one constructs the most general Lagrangian with the
same symmetries as the underlying theory QCD. In order to include the pseudoscalar
singlet field $\eta_0$, $SU(3)_L \times SU(3)_R$ chiral symmetry of conventional
ChPT is extended to $U(3)_L \times U(3)_R$ by treating the QCD vacuum angle $\theta$
as an external field which compensates the axial $U(1)$ anomaly \cite{B, L}.

The pseudoscalar meson nonet is summarized in a unitary matrix $U \in U(3)$ 
\begin{equation}
 U(\phi,\eta_0) = u^2 (\phi,\eta_0) = 
\exp \left(  i \frac{\sqrt{2}}{f_\pi} \phi + i \sqrt{\frac{2}{3}} \frac{\eta_0}{f_\pi}  \right) ,
\end{equation}
where $f_\pi \simeq 92.4$ MeV is the pion decay constant and $\phi$ contains
the Goldstone bosons $(\pi,K,\eta_8)$.
The ground state baryon octet $(N, \Lambda, \Sigma, \Xi)$ is also collected
in a $3 \times 3$ matrix and the effective Lagrangian is a function of
$U$, $B$ and their derivatives,
${\cal L}= {\cal L}( U, \partial U,\partial^2 U, 
          \ldots ,B,\partial B, \ldots,
           {\cal M},  \sfrac{\sqrt{6}}{f_\pi}\, \eta_0 + \theta )
$,
with ${\cal M}= \mbox{diag} (m_u, m_d, m_s)$ being the quark mass matrix and 
$\theta$ the QCD vacuum angle which appears in 
the gauge invariant combination with the singlet field $\eta_0$.

The $U(3)_L \times U(3)_R$ chiral effective Lagrangian of
the pseudoscalar meson nonet 
coupled to the ground state
baryon octet can be decomposed as
\begin{equation}
{\cal L} = {\cal L}_M + {\cal L}_{M B}
\end{equation}
with the mesonic piece up to  second chiral order \cite{BB}
\begin{equation}  \label{mes}
{\cal L}_M = 
- \frac{v_0}{f_\pi^2} \eta_0^2 + \frac{f_\pi^2}{4}  \langle u_{\mu}
u^{\mu} \rangle + \frac{f_\pi^2}{4} \langle \chi_+ \rangle + 
   i \frac{v_3}{f_\pi} \eta_0 \langle \chi_- \rangle ,
\end{equation}
where $\langle \ldots \rangle$ denotes the trace in flavor space.
The object
$u_{\mu} = i u^\dagger \nabla_{\mu} U u^\dagger$
contains the covariant derivative of the meson fields 
\begin{equation}
\nabla_{\mu} U  = \partial_\mu U + i e {\cal A}_\mu [Q, U]
\end{equation} 
with $Q = \frac{1}{3} \mbox{diag}(2,-1,-1)$
being the quark charge matrix and ${\cal A}_\mu$ the photon field.
Explicit chiral symmetry breaking is induced via the quark mass matrix
${\cal M}$
which enters in the combinations $\chi_\pm = 2 B_0 (u^\dagger {\cal M}
u^\dagger \pm u {\cal M} u  )$
with $B_0 = - \langle  0 | \bar{q} q | 0\rangle/ f_\pi^2$ the order
parameter of spontaneous symmetry violation.

The second and third term of Eq.\ (\ref{mes}) appear already
in conventional $SU(3)$ ChPT whereas the first and fourth one are due
to the axial $U(1)$ anomaly. The first one is the mass term of the
singlet field $\eta_0$ which remains in the chiral limit of vanishing
quark masses. The coefficient $v_0 $ is a parameter
not fixed by chiral symmetry and in the large $N_c$ limit it is
proportional to the topological susceptibility of Gluodynamics. The
fourth term contributes to $\eta_8$-$\eta_0$
mixing.

The meson-baryon interactions are collected in ${\cal L}_{\phi B}$ 
which reads at lowest chiral order \cite{B}
\begin{eqnarray}  \label{bar}
{\cal L}_{M B}^{(1)} &=& i \langle \bar{B} \gamma_{\mu} [D^{\mu},B] \rangle 
 - \mnod \langle \bar{B}B \rangle + i u_1 \frac{\eta_0^2}{f_\pi^2}
 \Big(  \langle [D^{\mu},\bar{B}]  \gamma_{\mu}  B \rangle - 
 \langle \bar{B}  \gamma_{\mu}  [D^{\mu},B] \rangle \Big) \no \\
&-&   \frac{1}{2} D \langle \bar{B} \gamma_{\mu}
 \gamma_5 \{u^{\mu},B\} \rangle  
- \frac{1}{2} F \langle \bar{B} \gamma_{\mu} \gamma_5 [u^{\mu},B] \rangle 
- \frac{1}{2} D_s  \langle \bar{B} \gamma_{\mu} \gamma_5 B \rangle 
  \langle u^{\mu} \rangle ,
\end{eqnarray}
where only the terms that are necessary for the present
calculation  are kept and the superscript denotes the chiral order.
$\mnod$ is the common baryon octet mass in the
chiral limit, $D,F, D_s$ are the axial vector couplings of the mesons
to the baryons and $D_\mu$ is the covariant derivative of the baryon fields.
Terms from ${\cal L}_{M B}^{(2)}$ are also included
in the calculation, but are not shown here for brevity, {\it cf.} \cite{BMW} for details.


\subsection{The coupled channel approach}

First, the $s$ wave interaction kernel $V$ of meson-baryon scattering
is extracted from the contact and $s$-channel Born 
terms.
Unitarity imposes a restriction on the $T$-matrix
\begin{equation} \label{unitarity}
T^{-1} =  V^{-1} + G ,
\end{equation}
where $G$ is the scalar meson baryon loop integral
\begin{equation}
G(q^2) = \int \frac{d^d l}{(2 \pi)^d} \frac{i}{[ (q-l)^2 - M_B^2 + i \epsilon]
   [ l^2 - m_\phi^2 + i \epsilon]} 
\end{equation}
with  $\mbox{Im} \, G = \mbox{Im} \, T^{-1}$,  and we have approximated the remaining 
real part in Eq. (\ref{unitarity}) by $V^{-1}$.
Matrix inversion of Eq. (\ref{unitarity}) yields the Bethe-Salpeter equation
\begin{equation}
T = [1+V \cdot G]^{-1} \cdot V 
\end{equation}
which is equivalent to the summation of a bubble chain.

This approach is readily extended to electroproduction of mesons on baryons.
The electric dipole amplitude $B_0^+$ and the longitudinal $s$ wave $C_0^+$
at the tree level
are derived from the contact and Born terms of meson electroproduction 
and inserted into the meson-baryon bubble chain, in order
to obtain the {\it full} electric dipole amplitude $E_0^+$ and longitudinal
$s$ wave $L_0^+$, respectively,
\begin{equation} \label{full}
E_0^+ = [1+V \cdot G]^{-1} \cdot B_0^+ , \qquad
L_0^+ = [1+V \cdot G]^{-1} \cdot C_0^+  ,
\end{equation}
which is illustrated in Figure~1.
\begin{figure}[ht]
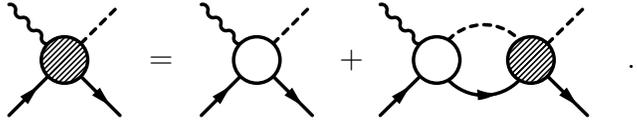

\[
\parbox{2cm}{\centering\includegraphics[scale=0.7]{feyn.30}}
=
\parbox{2cm}{\centering\includegraphics[scale=0.7]{feyn.31}}
+
\parbox{3cm}{\centering\includegraphics[scale=0.7]{feyn.32}} \quad .
\]
\caption{Shown is the electroproduction of mesons on baryons. The empty circle
denotes electroproduction at the tree level, whereas the full circles are the full
meson-baryon scattering and electroproduction amplitudes.
Wavy, dashed and solid lines represent the photon, mesons, and baryons, respectively.}
\label{fulldia}
\end{figure}

The $s$ wave total cross section for the electroproduction of mesons on the nucleon is given by
\begin{equation}
\sigma_{tot} = 8 \pi \frac{\sqrt{s} |\mbox{\boldmath$q$}|}{s -M_N^2}
\left(  |E_{0+}|^2  + \epsilon_L |L_{0+}|^2 \right) ,
\end{equation}
with $\mbox{\boldmath$q$}$ the three-momentum  of the meson in the center-of-mass frame and 
$\epsilon_L = -4 \epsilon s k^2 (s-M_N^2+k^2)^{-2}$ where $\epsilon$ and $k^2$ are the
virtual photon polarization and momentum transfer, respectively.

\section{Results}

In this section we will present the results of our calculation.
Some of the parameters in our approach are constrained by the octet baryon
masses and the $\pi N$ $\sigma$-term, while others are estimated assuming resonance
saturation. The remaining parameters are determined by
performing a global fit to available data for 
meson-proton and photon-proton reactions. 
This allows us to give  predictions for further processes
such as the cross sections for $\pi^- p \rightarrow \eta' n$
and $\eta$ electroproduction.

\subsection{Fit to the data}

We have performed a global fit to a large amount of data, consisting of
meson-proton and photon-proton reactions for values of $\sqrt{s}$
between 1.5 and 2.0 GeV.
Our cross sections include only $s$ wave
contributions, which are dominant in this energy range
for most processes. 

\begin{figure}[ht]
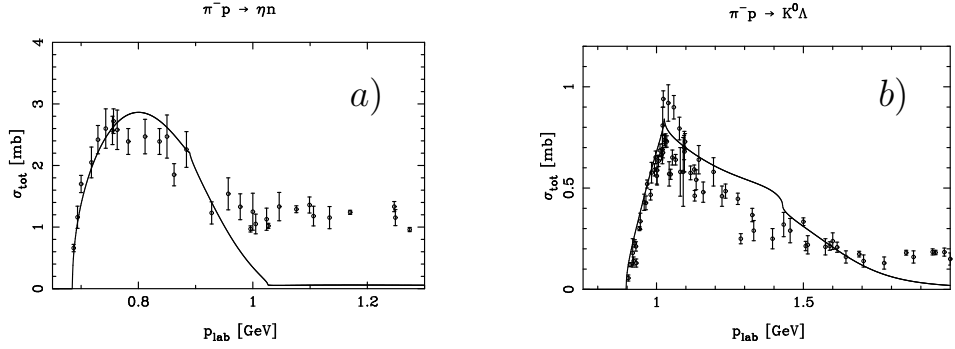

\centering
\begin{picture}(300,140)
\put(0,10){\makebox(100,120){\epsfig{file=fig5a.ps,width=4.5cm,angle=-90}}}
\put(200,10){\makebox(100,120){\epsfig{file=fig5b.ps,width=4.5cm,angle=-90}}}
\put(100,95){\large $a)$}
\put(300,95){\large $b)$}
\end{picture}
\caption{Total cross sections for pion-proton collision processes. The data
are taken from \cite{bald}.}
\label{fig:pip}
\end{figure}

Two sample pion-proton cross sections are shown in Fig.\ \ref{fig:pip}:
Fig.\ \ref{fig:pip}.a  shows
the cross section for the reaction $\pi^- p \rightarrow \eta n$. 
For momenta  $p_{\mbox{\scriptsize lab}}$   below 1 GeV the
cross section is dominated by the $s$ wave resonance $S_{11}(1535)$,
which is nicely reproduced in our calculation. 
In Fig.\ \ref{fig:pip}.b the results
for the reaction $\pi^- p \rightarrow K^0 \Lambda$ are given. Again the
cross section is dominated by $s$ waves, but in this case
the main contribution stems from the $S_{11}(1650)$ which is accompanied
by a cusp effect
due to the opening of the $K \Sigma$ threshold.

\begin{figure}[hb]
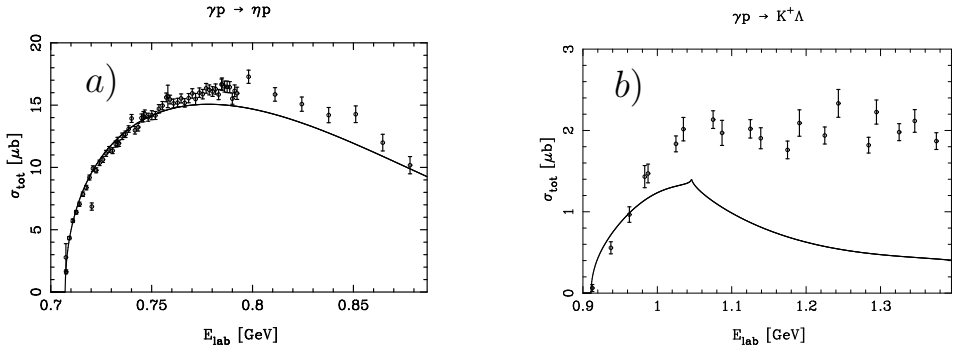

\centering
\begin{picture}(300,130)
\put(0,0){\makebox(100,120){\epsfig{file=fig6a.ps,width=4.5cm,angle=-90}}}
\put(200,0){\makebox(100,120){\epsfig{file=fig6b.ps,width=4.5cm,angle=-90}}}
\put(0,92){\large $a)$}
\put(200,90){\large $b)$}
\end{picture}
\caption{Shown are the total cross sections for $\eta$ and kaon photoproduction
off the proton. The data are taken from
\cite{mami, elsa, tran}.}
\label{fig:gamp}
\end{figure}

Let us now turn to the photoproduction cross sections which are shown
in Fig.\ \ref{fig:gamp}. We present in Fig.\ \ref{fig:gamp}.a the $\eta$
photoproduction data measured at MAMI \cite{mami} and ELSA \cite{elsa}
and our fitted result confirms 
the dominance of the $S_{11}(1535)$, which is responsible
for almost the entire cross section  in the low-energy region.
Results for the reaction $\gamma p \rightarrow K^+ \Lambda$ are given
in Fig.\ \ref{fig:gamp}.b, where at low
energies the cross section is dominated by the $S_{11}(1650)$,
while $p$ waves become important for $E_{\mbox{\scriptsize lab}}$
energies above 1.1 GeV. Again one observes a cusp effect due to the
$K \Sigma$ threshold.

\subsection{$\mbox{\boldmath$\eta'$}$ production}

\begin{figure}[ht]
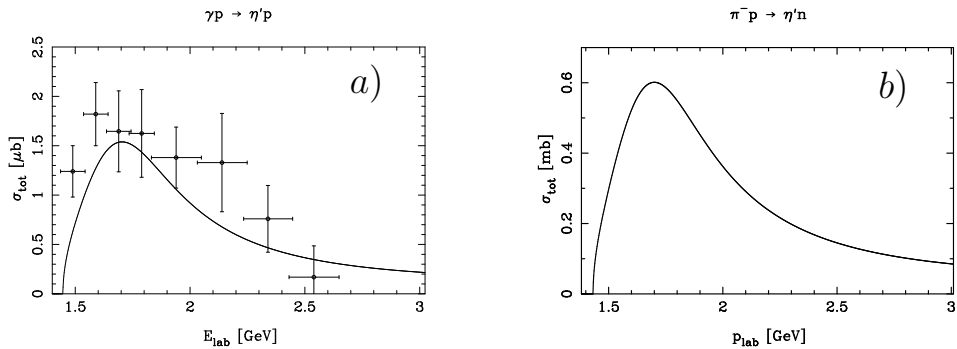

\centering
\begin{picture}(300,130)
\put(0,0){\makebox(100,120){\epsfig{file=fig7.ps,width=4.5cm,angle=-90}}}
\put(200,0){\makebox(100,120){\epsfig{file=fig5f.ps,width=4.5cm,angle=-90}}}
\put(100,92){\large $a)$}
\put(300,90){\large $b)$}
\end{picture}
\caption{In $a)$ the total cross section for $\eta'$-photoproduction
off the proton is given and the data are taken from \cite{Pl}. The total cross section
for pion-induced
$\eta'$-production is shown in $b)$.}
\label{fig:fig7}
\end{figure}

We show in Fig.\ \ref{fig:fig7}.a our result for the reaction
$\gamma p \rightarrow \eta' p$, which has been measured at ELSA \cite{Pl},
where the coherent contribution of two resonances, the $S_{11}(1897)$
and $P_{11} (1986)$, was observed.
Our formalism is capable of reproducing
the appearance of an $s$ wave resonance around 1.9 GeV.
Predictions  for the cross section of
$\pi^- p \rightarrow \eta' n$ are shown in Fig.\ \ref{fig:fig7}.b.

\subsection{Electroproduction of $\mbox{\boldmath$\eta$}$  mesons} 
\label{subsec:electro}

\begin{figure}
\centering
\begin{picture}(330,125)
\put(0,20){\makebox(100,120){\epsfig{file=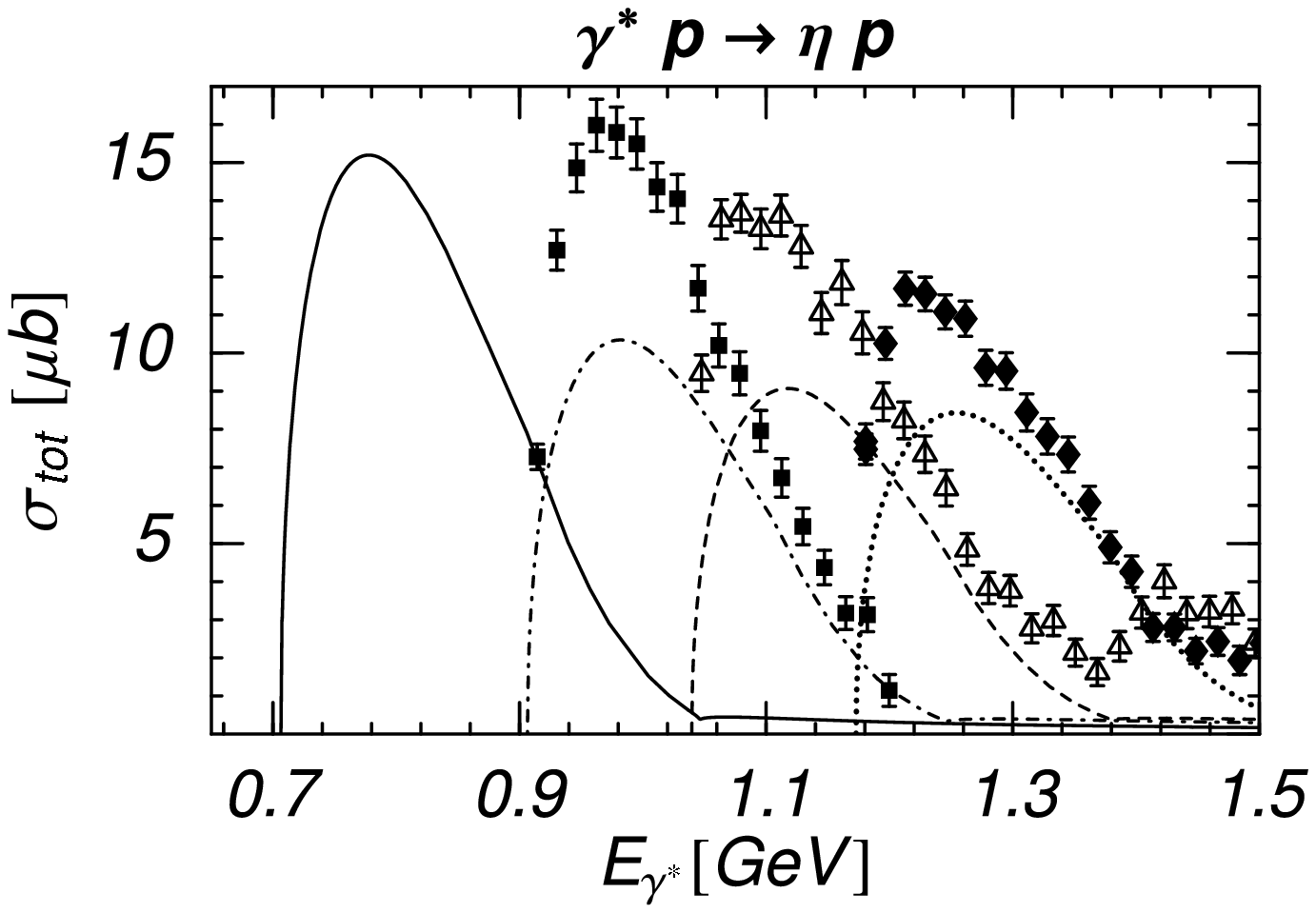,width=6.5cm}}}
\put(220,20){\makebox(100,120){\epsfig{file=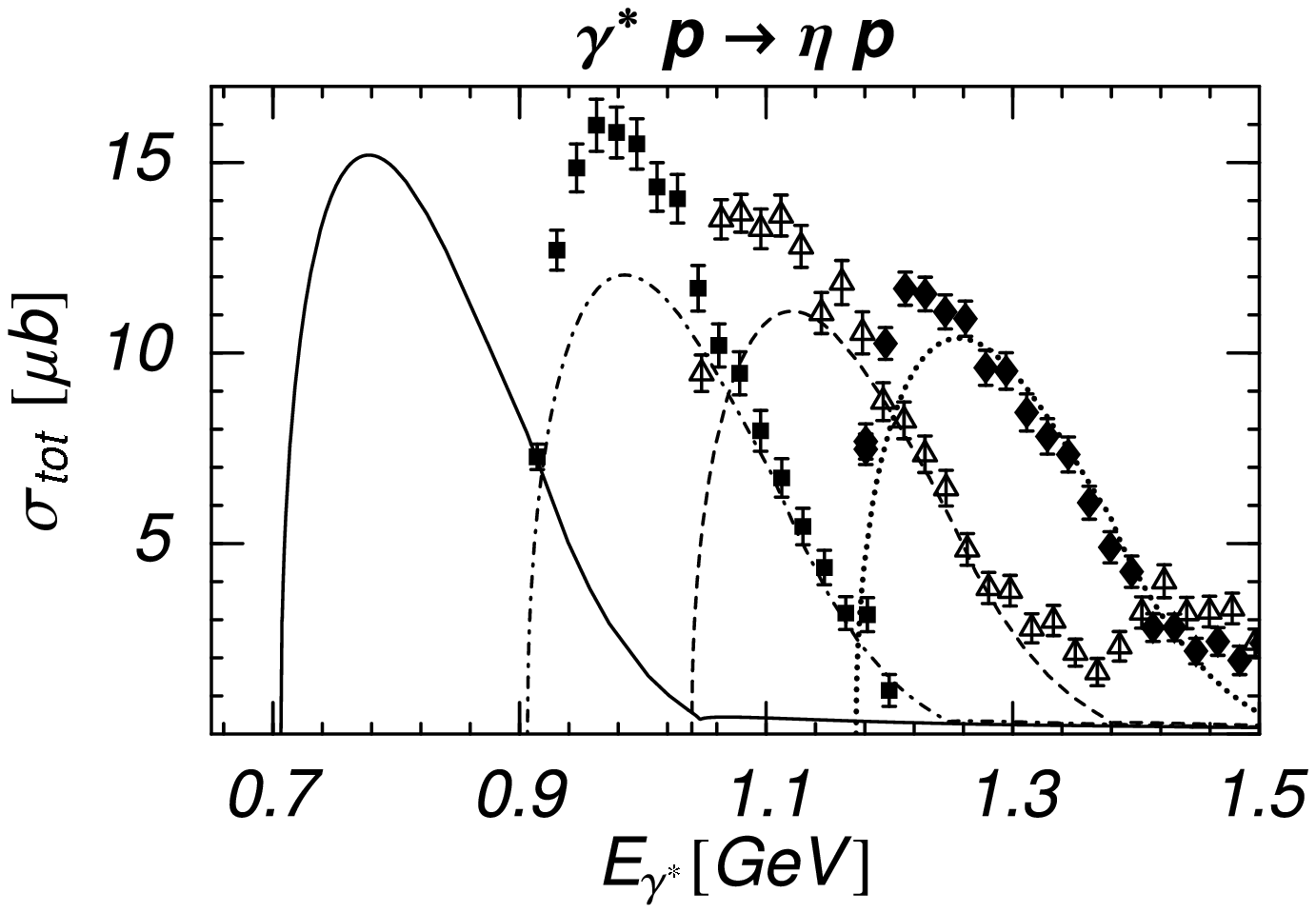,width=6.5cm}}}
\put(115,115){$a)$}
\put(332,115){$b)$}
\end{picture}
\caption{Cross sections for $\eta$ electroproduction on the proton
for various invariant momentum transfers $Q^2$
without $(a)$ and with $(b)$ transition form factors. The different lines refer
to the following values of $Q^2$: solid line: photoproduction ($Q^2=0$);
dash-dotted line and squares: $Q^2=0.375$ GeV$^2$; dashed line and
triangles: $Q^2=0.625$ GeV$^2$; dotted line and diamonds: $Q^2=0.875$ GeV$^2$.
The data with the same $Q^2$ values  
are taken from \cite{thomp}, and only statistical errors are shown.}
\label{fig:elp}
\end{figure}

We can furthermore compare our predictions for $\eta$ electroproduction with
available experimental data. We first present our predictions
for pointlike hadrons.
The $\eta$ electroproduction on the proton
has been measured in detail at CLAS at JLab \cite{thomp} and the data
is shown together with our results in Fig.\ \ref{fig:elp}.a.
The invariant momentum transfer $Q^2$ of the presented data
ranges from 0.375 to 0.875 GeV$^2$ and the applicability of our approach
to such high momentum transfers may be regarded as questionable.
Nevertheless, we should be able to capture qualitative features of
the $Q^2$ evolution of the cross section, in particular its
slow fall-off which is unusual and in sharp contrast, {\it e.g.}, to the fall
given by a nucleon dipole form factor. Our results, 
although showing less decrease with $Q^2$ than that of a simple
nucleon dipole form factor, produce a faster
reduction at low $Q^2$ than the experimental data, and then
flatten at higher momentum transfers.

The composite structure of
baryons and mesons will become increasingly
important with rising invariant momentum transfers $Q^2$.
It is therefore natural to include form factors in the initial 
electroproduction potentials $B_{0+}$ and $C_{0+}$, in order to account
for the electromagnetic structure of hadrons. One may be inclined
to multiply the potentials by an overall form factor which is the same for all
hadrons that interact with the photon (and thus the same 
for all channels). However, this procedure will also 
yield smaller cross sections for $\eta$ electroproduction 
leading to stronger disagreement with the data.
In this case, the inclusion of form factors which accounts for the
electromagnetic structure of the baryons and mesons seems to worsen
the situation.

In order to model the electromagnetic structure of $B_{0+}$ and $C_{0+}$,
we include  monopole form factors $(1+Q^2/M_\alpha^2)^{-1}$ 
with mass parameters $M_\alpha$ depending
on the outgoing channel $\alpha$. The results for $\eta$ 
electroproduction after including these form factors are given in 
Fig.\ \ref{fig:elp}.b.
Surprisingly, the cross sections for $\eta$ electroproduction are {\it increased}. 
This is due to the fact that
we employed different form factors for the participating channels and that
some of these channels may compensate each other in the final state
interactions. In contrast to common belief, the inclusion of form factors
can increase the cross section, {\it e.g.} in $\eta$ electroproduction, yielding
a $Q^2$ evolution closer to experiment and indicating a hard transition
form factor. The usual claim that the hard form factor is counterintuitive to
an interpretation of this state as a bound hadronic system 
is not justified, since within the model the photon couples directly to
one of the ground state octet baryons or  a meson 
and only after this initial 
reaction the produced meson forms a bound state with the baryon.

We do not 
expect our results for electroproduction to reproduce precisely
the experimental data. Nevertheless, the inclusion of simple form factors
for the electroproduction potentials is able to explain
qualitatively the slow
decrease of the $S_{11}(1535)$ photocoupling.

\subsection{Effects of the $\eta'$} \label{subsec:effe}

\begin{figure}[hb]
\centering
\begin{picture}(330,125)
\put(0,0){\makebox(100,120){\epsfig{file=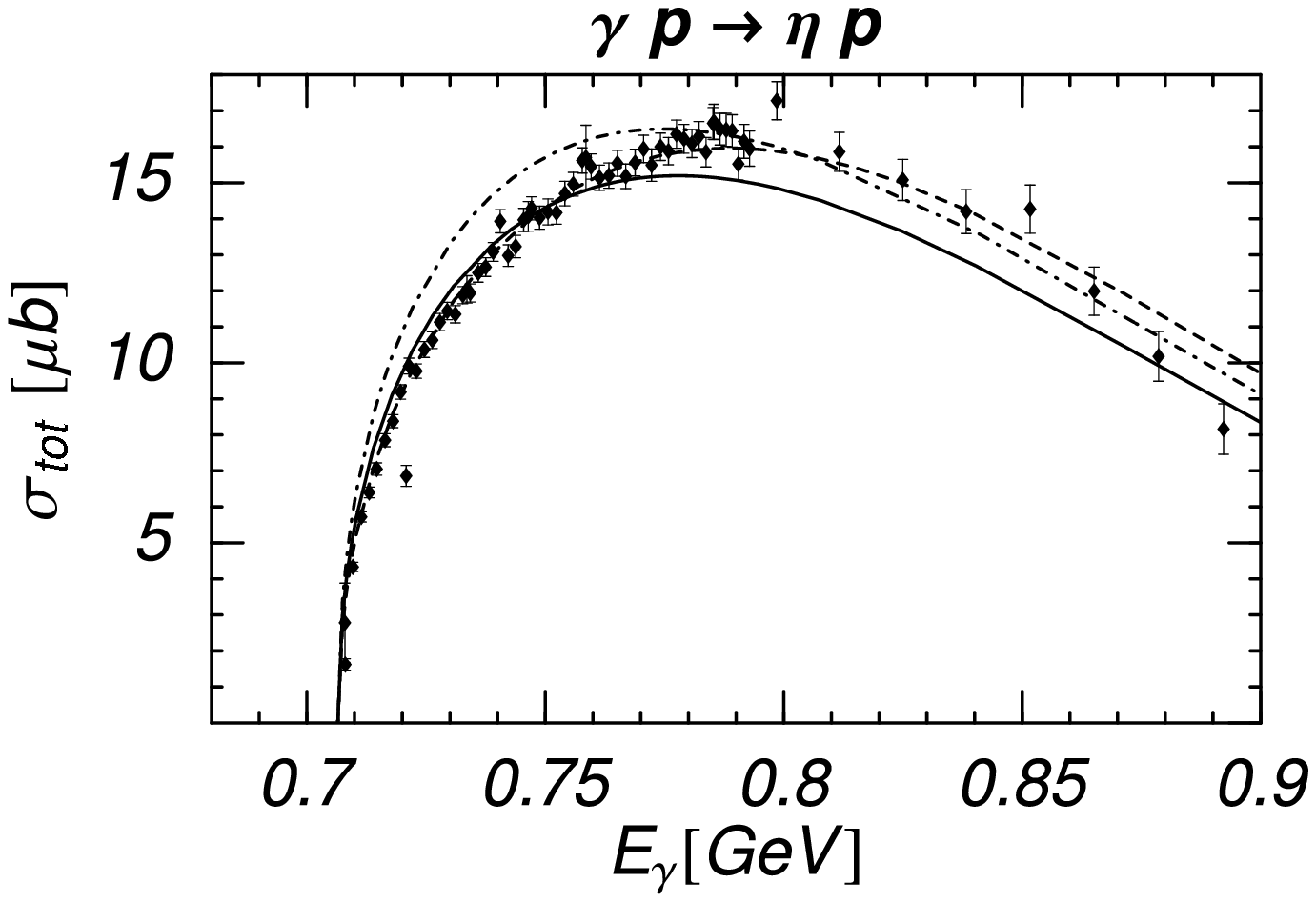,width=6.5cm}}}
\put(220,0){\makebox(110,120){\epsfig{file=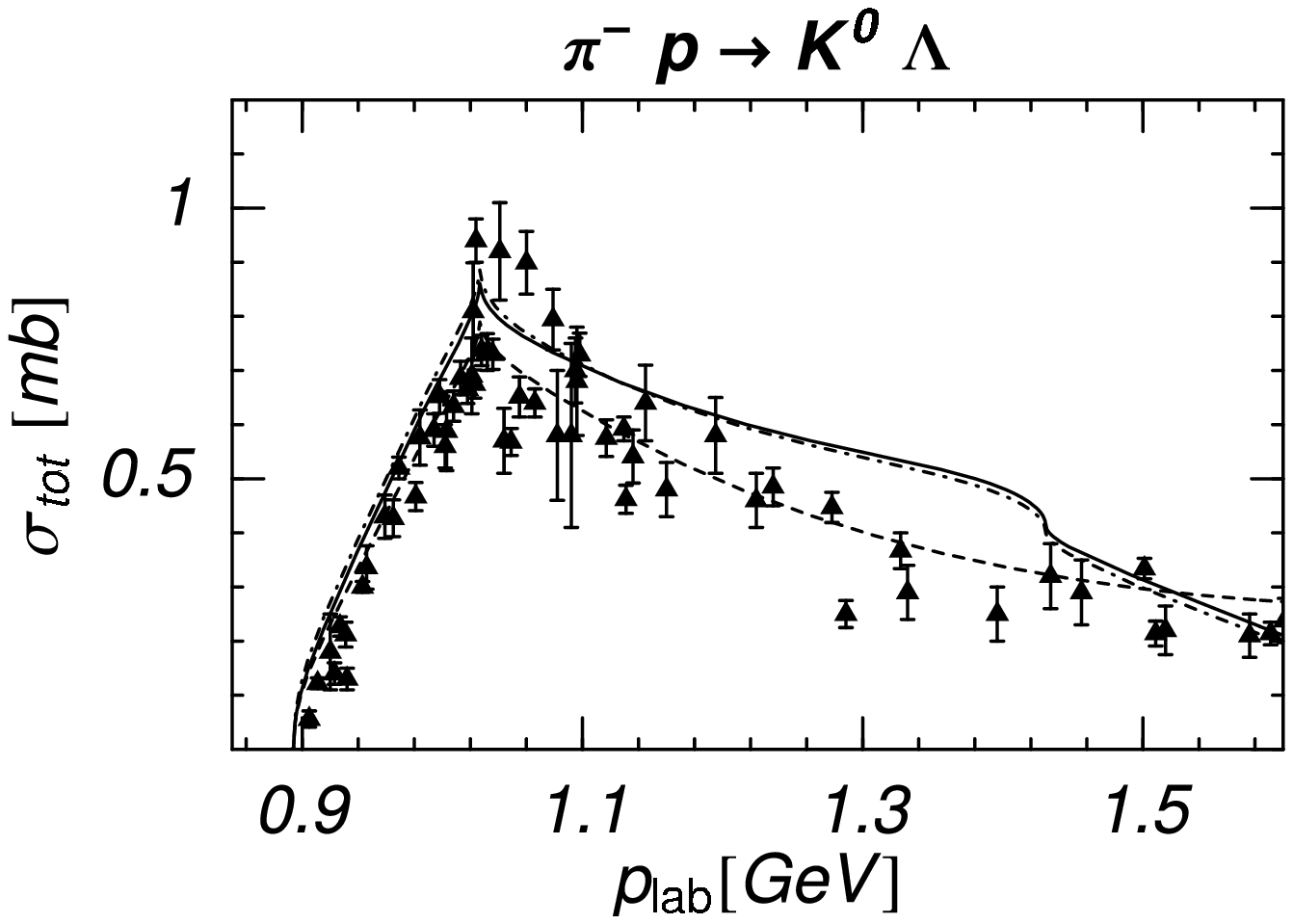,width=6.5cm}}}
\put(-5,95){$a)$}
\put(225,95){$b)$}
\end{picture}
\caption{Shown are the differences in the cross sections for
$\gamma p \rightarrow  \eta p $ 
and $\pi^- p \rightarrow K^0 \Lambda$ after neglecting $\eta$-$\eta'$ mixing 
and the $| \eta' N \rangle$ channel in the coupled channel formalism.
The solid line is the original result, the dash-dotted line is obtained for
vanishing $\eta$-$\eta'$ mixing, and the dashed line refers to the case without
the $| \eta' N \rangle$ channel.}
\label{fig:effe}
\end{figure}

In this section, we wish to investigate the effects of $\eta$-$\eta'$
mixing and the importance of the $| \eta' N \rangle$ virtual state in our
coupled channel formalism. This is done in a two-step procedure: first,
$\eta$-$\eta'$ mixing is turned off, and secondly, we eliminate the
$| \eta' N \rangle$ channel from the model. In both cases we do not
repeat the fit which would actually compensate most
of the changes. We are particularly interested in the reaction
$\pi^- p \rightarrow K^0 \Lambda$ which exhibits the most prominent
$\eta' N$ cusp of all the channels. Omission of both $\eta$-$\eta'$ mixing 
and the $| \eta' N \rangle$ channel do not lead to substantial
differences in the remaining reactions (where the $\eta'$ is not produced).
In Fig.\ \ref{fig:effe} we have chosen to present in addition to 
$\pi^- p \rightarrow K^0 \Lambda$ the photoproduction process
$\gamma p \rightarrow \eta p$, in order to give a measure for the changes
in the other channels.
For $\pi^- p \rightarrow K^0 \Lambda$ variation in $\eta$-$\eta'$ mixing has
almost no impact (dash-dotted line), like in most other channels in which
the $\eta$ is not
produced as a final particle. Eliminating the $| \eta' N \rangle$ channel 
makes the $\eta' N$ cusp disappear and lowers the cross section, bringing it 
to better agreement with the data (dashed line). It suggests that the
region around the $\eta' N$ cusp is
overemphasized within our model. This feature may change after the inclusion of
$p$ waves, since then a new overall fit to the different reaction channels
will lower the $s$ wave contribution reducing the absolute importance of
the cusp.
For the photoproduction of the $\eta$ on the proton $\eta$-$\eta'$ mixing 
plays a slightly more prominent role (dash-dotted line), as the $\eta$ is
produced in the final state. If the $| \eta' N \rangle$ channel is turned
off, the changes are again quite moderate (dashed line).
Overall we can conclude, that the results for the production of the
Goldstone bosons are not modified substantially after omitting the
$\eta'$ which is in accordance with
intuitive expectation, since the  $| \eta' N \rangle$ channel is much higher
in mass than the other channels. 
We can therefore confirm that it was justified in previous coupled channel
analyses to neglect $\eta$-$\eta'$ mixing and
treat the $\eta$ as a pure octet state, see e.g. \cite{KWW}.

\subsection*{Acknowledgement}
This work was supported by the Deutsche Forschungsgemeinschaft.

\end{document}